\documentclass{epl}
\usepackage{bm}
\usepackage{graphicx}
\newcommand{\tQ}{\tilde{Q}}
\newcommand{\tq}{\tilde{q}}

\title{Modeling planar degenerate wetting and anchoring in\\ nematic
liquid crystals}
\shorttitle{Planar degenerate wetting and anchoring}
\author{J.-B. Fournier\inst{1,2} \and P. Galatola\inst{1}}

\institute{
  \inst{1}Mati\`ere et Syst\`emes Complexes, UMR 7057 CNRS
\& Universit\'e Paris~7 - 2 place Jussieu, F-75251 Paris Cedex 05,
France\\
  \inst{2}Laboratoire de Physico-Chimie Th\'eorique, UMR 7083
  CNRS-ESPCI, 10 rue Vauquelin, F-75231 Paris Cedex 05, France
}
\pacs{61.30.Hn}{Liquid crystals: surface phenomena}
\pacs{61.30.Dk}{Continuum models and theories of liquid crystal
structure}
\pacs{64.70.Md}{Transitions in liquid crystals}

\begin{document}

\maketitle

\begin{abstract}
We propose a simple surface potential describing the planar degenerate
anchoring of nematic liquid crystals, i.e., the tendency of the
molecules to align parallel to one another along any direction parallel
to the surface. We show that, at lowest order in the tensorial Landau-de
Gennes order-parameter, fourth-order terms must be included.  We analyze
the anchoring and wetting properties of this surface potential.  In
the nematic phase, we find the desired degenerate planar anchoring, with
positive scalar order-parameter and some surface biaxiality. In the
isotropic phase, we find, in agreement with experiments, that the
wetting layer may exhibit a uniaxial ordering with negative scalar
order-parameter. For large enough anchoring strength, this negative
ordering transits towards the planar degenerate state.
\end{abstract}

\section{Introduction}

Nematic liquid crystals are non-polar liquids consisting of elongated
molecules aligned, on the average, parallel to a common direction
$\pm\bm{n}$~\cite{deGennes93}. The unit vector $\bm{n}$ is called the
\textit{director}. Exceptionally, in the case of ``negative order,'' the
molecules may be distributed perpendicularly to $\bm{n}$. The nematic
order is quantified by a symmetric traceless tensor~$\mathsf{Q}$, of
Cartesian components $Q_{ij}$ ($i,j=1,2,3$).  Microscopically,
calling $\bm{m}$ a unit vector parallel to the long axis of the
molecules, the tensor order-parameter can be defined as $Q_{ij}=\langle
m_im_j\rangle-\frac{1}{3}\delta_{ij}$, where the brackets denote
statistical averaging and $\delta_{ij}$ is the Kronecker symbol. For
uniaxial ordering, which corresponds to the preferred state of the nematic
in the bulk, two of the eigenvalues of $\mathsf{Q}$ are
equal; in this case $Q_{ij}$ can be written as
$S(n_in_j-\frac{1}{3}\delta_{ij})$, where $-\frac{1}{2}\le S\le 1$ is
the scalar order-parameter. Note that $S=0$
corresponds to the isotropic phase (randomly oriented molecules),
occurring at temperatures above the nematic--isotropic phase transition;
$S>0$ corresponds to the ordinary nematic, in which the molecules are
aligned on the average parallel to~$\bm{n}$, while $S<0$ corresponds to
the ``negative'' case, in which the molecules are on the average
orthogonal to~$\bm{n}$.

The alignment of a nematic liquid crystal by a bounding interface is of
considerable interest both for fundamental and technological reasons. In
the most common situations, this ``anchoring'' effect is well described
by a surface free energy depending only on the director $\bm{n}$ at the
interface\cite{deGennes93}. Three typical situations are: \textit{i})
\textit{homeotropic} anchoring, where the preferred, or ``easy'',
average orientation corresponds to $\bm{n}$ normal to the interface,
\textit{ii}) \textit{planar} anchoring, where the preferred average
orientation corresponds to $\bm{n}$ lying in one particular direction
parallel to the interface, and \textit{iii}) \textit{planar degenerate}
anchoring, where all the planar orientations for $\bm{n}$ are equivalent
easy directions. The latter situation is often realized with isotropic
bounding fluids, e.g., glycerin~\cite{Lavrentovich90} or
gallium~\cite{warenghem84}; degenerate anchorings can also be
obtained with polymer-grafted surfaces\cite{Ou00}.

In general, surface interactions also perturb the \textit{degree} of
nematic ordering, e.g., the value of~$S$. Above the bulk
nematic--isotropic transition temperature, this effect may induce the
\textit{wetting} of the surface by a nematic layer with $S\ne0$, as
predicted by Sheng~\cite{PingSheng}. This effect, long ago
observed by Miyano~\cite{Miyano79}, was also recently detected by a
surface force apparatus measuring the related capillary
condensation~\cite{Kocevar01}.  Although nematics are uniaxial in the
bulk, the wetting layer may exhibit \textit{biaxiality} due to the lower
symmetry near the surface~\cite{Sluckin85}. In the biaxial case, the
symmetric and traceless $Q_{ij}$ has its three eigenvalues $\lambda_i$
($i=1,2,3$) all different. A nematic director, $\bm{n}$, can still be
defined as the eigenvector corresponding to the eigenvalue of largest
modulus, say $\lambda_3$. By analogy with the uniaxial case, one can
then define the scalar order-parameter by $\lambda_3=\frac{2}{3}S$; the
amount of biaxiality, $0\le B\le1$, is then defined by
$\lambda_1=-\frac{1}{3}S(1-B)$ and $\lambda_2=-\frac{1}{3}S(1+B)$.

The aim of the present paper is to model the \textit{planar degenerate
anchoring} of nematic liquid crystals, using the tensorial Landau-de
Gennes formalism. To our knowledge, this has not been done before.
Actually, there is some confusion in the literature, since a number of
papers dealing with ``degenerate planar anchoring'' describe in fact
homeotropic anchoring with negative
$S$~\cite{Sluckin85,Sen87,Seidin97,Huber05}. The confusion arises from
the fact that the molecules themselves are planar and randomly oriented
in this homeotropic situation. Again, conventional planar degenerate
anchoring describes the tendency of a substrate to favor the
\textit{common alignment of the molecules along any direction parallel
to its surface}~\cite{deGennes93}. This is the very situation that we
consider in this paper.

\section{Theory}

Let us recall some of the existing tensorial anchoring models.  The
simplest quadratic surface free energy favoring a given
\textit{non-degenerate} nematic ordering $\mathsf{Q}^{(0)}$, i.e., a
well-defined director and a well-defined scalar and, possibly, biaxial
order-parameter, was proposed by Nobili and Durand~\cite{Nobili92}:
\begin{equation} 
f_\mathrm{ND}=\frac{1}{2}W\left(Q_{ij}-Q^{(0)}_{ij}
\right) \left(Q_{ij}-Q^{(0)}_{ij} \right).  
\end{equation} 
Here, and throughout, summation over repeated indices is implicit.
Because $f_\mathrm{ND}$ is proportional to the sum of the squares of the
components of $\mathsf{Q}-\mathsf{Q}^{(0)}$, it obviously has, for
$W>0$, a unique minimum for $\mathsf{Q}=\mathsf{Q}^{(0)}$. In this
simple situation, the surface energy \textit{itself} possesses a
well-defined minimum. In principle, this is not necessary: the
only requirement is that the nematic free energy as a whole,
i.e., the sum of the surface and the bulk free energies, be bounded from
below. For instance, early descriptions have dealt with only a
\textit{linear} surface free energy $\nu_i\,Q_{ij}\nu_j$, where
$\bm{\nu}$ is the normal to the interface~\cite{PingSheng}. Actually, it
is not clear at which order the surface free energy expansion should be
truncated. In principle, for any truncation, one should verify that the
higher-order terms are indeed negligible (this is especially critical in
the case of large surface couplings). Another difficulty, when dealing
with truncated surface free energies not bounded from below, is that one
cannot determine which kind of anchoring they favor without including
bulk terms and solving the whole problem. For these reasons, we shall
concentrate in this paper on surface free energies that possess
well-defined minima by themselves.

The most general \textit{quadratic} form describing isotropic substrates
is~\cite{Sen87,Sluckin86}
\begin{equation}
f_2=W_{11}\, \nu_i\, Q_{ij}\, \nu_j +W_{21}\,Q_{ij}\,Q_{ij}
+W_{22}\left(\nu_i\, Q_{ij}\, \nu_j\right)^2
+W_{23}\,\nu_i\, Q_{ij}\,Q_{jk}\,\nu_k.
\end{equation}
It contains all the scalars up to quadratic order made with $\mathsf{Q}$
and $\bm{\nu}$ (the surface normal). Minimizing $f_2$ with respect to
the five components of the most general symmetric traceless tensor
yields, in the basis where the $z$-axis is along $\bm{\nu}$, a unique
extremum~\cite{det}, uniaxial, at
$\mathsf{Q}^{(0)}=\mathrm{diag}(\lambda,\lambda,-2\lambda)$, where
$\lambda=W_{11}/(4W_{22}+4W_{23}+6W_{21})$. This corresponds to a
\textit{homeotropic} anchoring ($\bm{n}=\bm{\nu}$) with either $S>0$ or
$S<0$ depending on the sign of $\lambda$.

\begin{figure}
\centerline{\includegraphics[width=.5\textwidth]{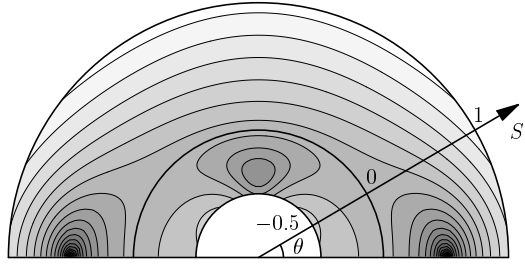}}
\caption{Polar contour plot of the surface potential~(\protect\ref{eq:f4}), with
$W_1=W_2$ and $S_0=0.5$, for a \textit{uniaxial} surface
order-parameter $Q_{ij}=S(n_i\,n_j-\frac{1}{3}\delta_{ij})$, where
$\bm{n}=(\cos\theta\cos\phi,\cos\theta\sin\phi,\sin\theta)$. The angle $\theta$ is the inclination of the director
with respect to the substrate and the radial coordinate corresponds to
the scalar order-parameter $-\frac{1}{2}\le S\le1$. The energy does not
depend on $\phi$. Darker shadings corresponds to lower energies. The
absolute minimum is attained for degenerate planar anchoring: $\theta=0$
(or $\pi$) and $S=S_0$; the apparent homeotropic minimum for
$\theta=\pi/2$ and $S<0$ is actually a saddle-point with respect to the
biaxial degree of freedom (outside the plane of this figure).}
\label{fig:polar} \end{figure}

To obtain a planar degenerate uniaxial minimum, it is necessary to
include terms of order higher than two. We shall go up to fourth order
to have a surface free energy bounded from below. Note that this is also
the lowest order for the bulk free energy to correctly describe the
nematic--isotropic phase transition. Our aim is to obtain a degenerate
manifold of minima for
$Q_{ij}=S_0\left(n_i\,n_j-\frac{1}{3}\delta_{ij}\right)$, where $\bm{n}$
is an arbitrary unit vector parallel to the substrate, and where $S_0>0$
is fixed. To this purpose, we define 
\begin{equation}
\tQ_{ij}=Q_{ij}+\frac{1}{3}S_0\,\delta_{ij},
\end{equation}
and we look for the constraints ensuring that $\tQ_{ij}$ takes the form
$S_0\,n_i\,n_j$, where $\bm{n}$ is parallel to the substrate. The two
following conditions are necessary and sufficient~\cite{dem}:
\textit{i}) $\tQ_{ij}$ must coincide with its projection on the
substrate $\tQ_{ij}^\perp=P_{ik}\,\tQ_{k\ell}\,P_{\ell j}$, where
$P_{ij}=\delta_{ij}-\nu_i\nu_j$, and \textit{ii}) the trace of the
square of $\tQ_{ij}$ must be equal to ${S_0}^2$, that is
$\tQ_{ij}\,\tQ_{ij}={S_0}^2$. The simplest surface free energy being
minimum if, and only if, these two conditions are satisfied
is 
\begin{equation}
\label{eq:f4}
f_4= W_1
\left(\tQ_{ij}-\tQ_{ij}^\perp\right)\left(\tQ_{ij}-\tQ_{ij}^\perp\right)
+W_2\left(\tQ_{ij}\,\tQ_{ij}-{S_0}^2\right)^2,
\end{equation}
with $W_1>0$ and $W_2>0$. This surface potential is of fourth-order in
$\mathsf{Q}$ and covariant. It is not, however, the most general
fourth-order expansion giving rise to a planar degenerate anchoring: one
can add, for instance, a contribution proportional to the square of the
first term and several other similar, but linearly independent, terms.
In the basis where the $z$-axis is along the surface normal $\bm{\nu}$, the
above surface potential takes the explicit form
\begin{eqnarray}
f_4&=&W_1\left[2\left(Q_{xz}^2+Q_{yz}^2\right) + 
  \left(Q_{xx} +Q_{yy} - \frac{S_0}{3} \right)^2\right]\nonumber\\
&+&4\,W_2\left(Q_{xx}^2 + Q_{yy}^2 + Q_{xx}\, Q_{yy}+ Q_{xy}^2 +
Q_{xz}^2 + Q_{yz}^2 - \frac{{S_0}^2}{3} \right)^2,
\end{eqnarray}
where we have taken into account that $\mathsf{Q}$ is symmetric and
traceless ($Q_{zz}=-Q_{xx}-Q_{yy}$).
As shown in fig.~\ref{fig:polar}, the surface potential $f_4$ possesses
a degenerate absolute minimum for uniaxial order-parameters with planar
director and $S=S_0>0$. No other minima are present: the apparent
homeotropic minimum with negative scalar order-parameter visible in
fig.~\ref{fig:polar} is actually a saddle point with respect to biaxial
perturbations.

\section{Wetting and anchoring properties}
To test the physical properties of the surface potential~(\ref{eq:f4}),
we consider a semi-infinite sample filled, for $z>0$,  with a liquid crystal
either in the nematic or in the isotropic phase. For the bulk free
energy density we take the fourth-order Landau-de Gennes
expansion~\cite{deGennes93}:
\begin{equation}
f_V=\frac{1}{2}a\left(T-T^\star\right)Q_{ij}\,Q_{ij}
-\frac{1}{3}b\,Q_{ij}Q_{jk}\,Q_{ki}
+\frac{1}{4}c\,\left(Q_{ij}\,Q_{ij}\right)^2
+\frac{1}{2}L\,Q_{ij,k}\,Q_{ij,k}\,,
\end{equation}
where the comma indicates spatial derivation. The coefficients $a$, $b$,
$c$ are assumed positive, $L>0$ is an elastic constant (in the one
constant approximation), $T$ is the temperature, and $T^\star$ is the
supercooling temperature of the isotropic phase. We define the normalized
temperature $\tau=27acb^{-2}(T-T^\star)$ and the rescaled order-parameter $q_{ij}=\frac{3}{2}\sqrt{6}\,b^{-1}c\,Q_{ij}$. The first-order
nematic--isotropic transition occurs for $\tau=1$. In the nematic phase
$(\tau<1)$, the reduced scalar order-parameter is
$s_b(\tau)=\frac{1}{4}\sqrt{3/2}(3+\sqrt{9-8\tau})$. We rescale the lengths
with respect to the correlation length at the transition
$\xi=b^{-1}\sqrt{27cL}$ and the energies with respect to
$F_0=2b[L/(3c)]^{3/2}$. The dimensionless bulk and surface free energy
densities become
$\frac{1}{2}\tau\,q_{ij}q_{ij}-\sqrt{6}\,q_{ij}q_{jk}\,q_{ki}
+\frac{1}{2}(q_{ij}q_{ij})^2+\frac{1}{2}q_{ij,k}q_{ij,k}$ and
$w_1(\tq_{ij}-\tq_{ij}^\perp) (\tq_{ij}-\tq_{ij}^\perp)
+w_2(\tq_{ij}\,\tq_{ij}-{s_0}^2)^2$, respectively, with the normalized
anchoring parameters $w_1=3b^{-1}(3c/L)^{1/2}\,W_1$,
$w_2=\frac{2}{3}b(3Lc^3)^{-1/2}\,W_2$ and
$s_0=\frac{3}{2}\sqrt{6}\,b^{-1}c\,S_0$. We look for equilibrium
order-parameter profiles depending only on $z$ and possessing
a symmetry plane orthogonal to the substrate, which we take as the
$(y,z)$ plane. Therefore, we set
\begin{equation}
\mathsf{q}(z)=\pmatrix{
-\alpha(z)\!-\!\beta(z)&0&0\cr
0&\alpha(z)&\gamma(z)\cr
0&\gamma(z)&\beta(z)}.
\end{equation}
The minimization problem defined by $f_V$ and $f_4$ yields the
Euler-Lagrange equations in the bulk:
\begin{eqnarray}
\frac{d^2\alpha}{dz^2}&=&
\tau\alpha-\sqrt{6}\left(\alpha^2-2\beta^2+\gamma^2-2\alpha\beta\right)
+4\alpha\left(\alpha^2+\beta^2+\gamma^2+\alpha\beta\right),\\
\frac{d^2\beta}{dz^2}&=&
\tau\beta-\sqrt{6}\left(\beta^2-2\alpha^2+\gamma^2-2\alpha\beta\right)
+4\beta\left(\alpha^2+\beta^2+\gamma^2+\alpha\beta\right),\\
\frac{d^2\gamma}{dz^2}&=&
\gamma\left[\tau-3\sqrt{6}\left(\alpha+\beta\right)
+4\left(\alpha^2+\beta^2+\gamma^2+\alpha\beta\right)\right],
\end{eqnarray}
with the boundary conditions on the substrate, i.e., at $z=0$,
\begin{eqnarray}
\frac{d\alpha}{dz}&=&
\frac{8}{3}w_2\,\alpha\left[3\left(\alpha^2+\beta^2+\gamma^2+\alpha\beta\right)-s_0^2\right]-\frac{2}{9}w_1\left(3\beta+s_0\right),\\
\frac{d\beta}{dz}&=&
\frac{8}{3}w_2\,\beta\left[3\left(\alpha^2+\beta^2+\gamma^2+\alpha\,\beta\right)-s_0^2\right]+\frac{4}{9}w_1\left(3\beta+s_0\right),\\
\frac{d\gamma}{dz}&=&\gamma\left[
\frac{8}{3}w_2\left[3\left(\alpha^2+\beta^2+\gamma^2+\alpha\,\beta\right)-s_0^2\right]+\frac{6}{3}w_1\right],
\end{eqnarray}
and for $z\to\infty$, $d\alpha/dz=d\beta/dz=d\gamma/dz=0$.

We solve numerically the above boundary value problem by a finite
difference algorithm with deferred correction and Newton
iteration~\cite{bvp}. When multiple solutions are present, we select the
one with the lowest total free energy. For the sake of simplicity, we
set $w_1=w_2=w$ and, for a given value of $s_0$, we explore the behavior
of the order-parameter profile for different values of $\tau$ and $w$.
We find that the eigenvectors of the order-parameter solutions are 
always parallel to the $x$, $y$ and $z$ axes, which corresponds to
$\gamma(z)=0$.

In the nematic phase ($\tau<1$), we find the expected planar anchoring,
with, however, a slight biaxiality in a surface layer of thickness
$\simeq\xi$. The nematic director $\bm{n}$ is parallel
to the surface and uniform along $z$, the scalar order-parameter $s>0$
relaxes over $\simeq\xi$ to the bulk value $s_b$ starting from a surface
value~$s_s$ intermediate between $s_0$ and $s_b$. The amount of
biaxiality at the surface, $B_s$, vanishes for $s_0=s_b$ and grows with
$|s_0-s_b|$; it obviously vanishes for $w=0$, but also for $w\to\infty$,
presenting a maximum at a finite value of~$w$.

In the isotropic phase ($\tau>1$), the order-parameter vanishes in the
bulk, but a wetting nematic layer appears at the surface. For
large values of $w$, the ordering at the surface is planar and biaxial,
as in the nematic phase. We call this state $P^\dag$ (the dagger
indicating biaxiality). As before, the amount of biaxiality tends to
zero as $w\to\infty$. Conversely, for small values of $w$, the ordering
at the surface is homeotropic and uniaxial, with a \textit{negative}
scalar order-parameter. We call this state $H_n$. In between, a surface
transition occurs. 

\begin{figure}
\centerline{\includegraphics[width=.4\textwidth]{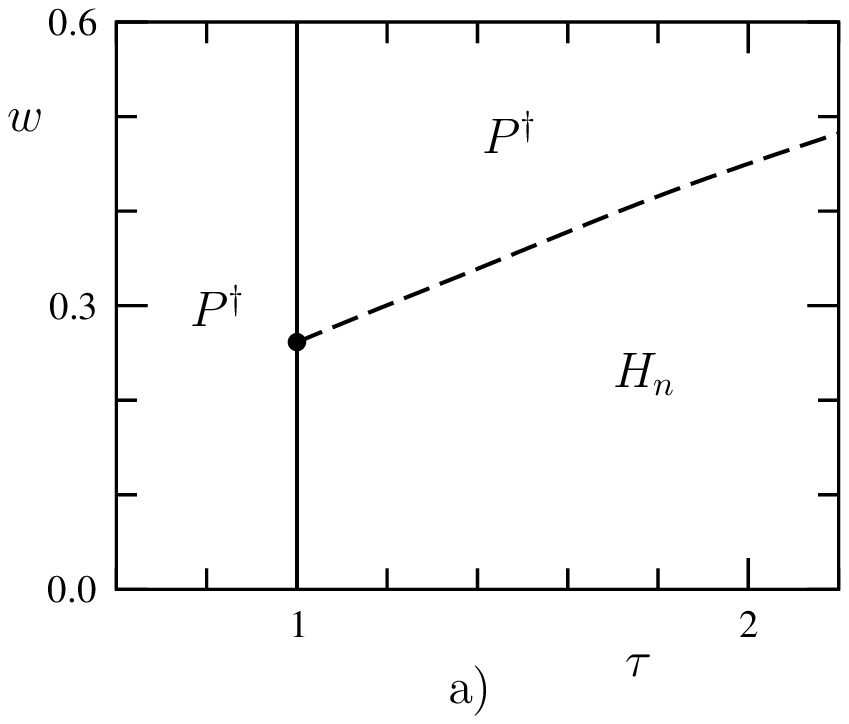}\hfil
\includegraphics[width=.4\textwidth]{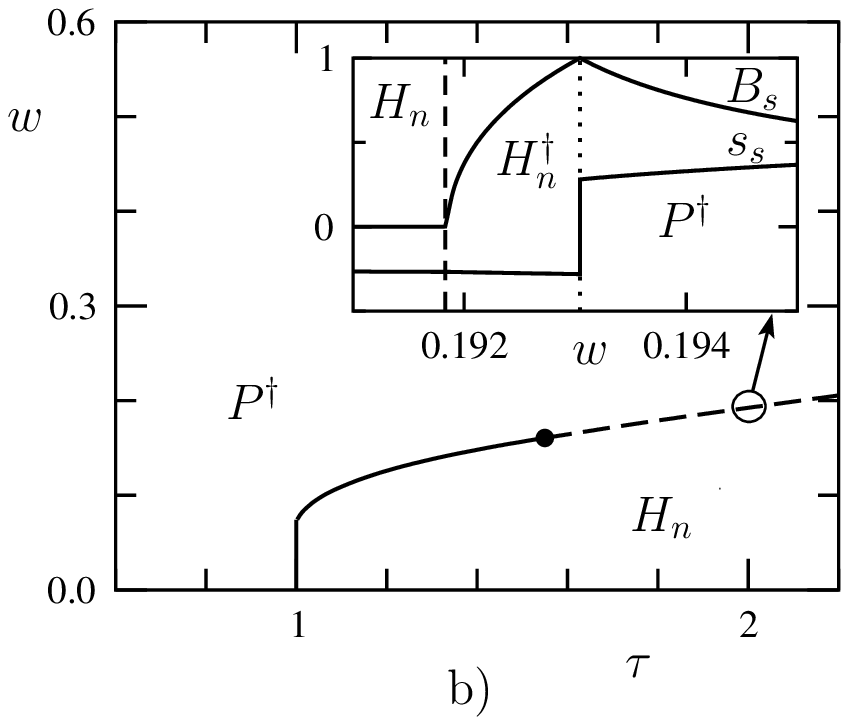} }
\caption{a) Phase diagram of the surface state in the plane of the
reduced temperature $\tau$ and of the reduced anchoring
strength~$w=w_1=w_2$, for the preferred surface scalar order-parameter
$s_0=1<s_b(1)$. The states $P^\dag$ and $H_n$ correspond to a planar
biaxial ordering with positive scalar order-parameter and to a
homeotropic uniaxial ordering with negative scalar order-parameter,
respectively.  Full (resp.\ dashed) lines indicate first
(resp.\ second) order transitions. The second-order line terminates at a
critical end-point.  b) Same as a) but for $s_0=1.5>s_b(1)$. The dot
indicates a tricritical point. Inset: Evolution of the reduced surface
scalar order-parameter, $s_s$, and of the amount of biaxiality, $B_s$,
as a function of $w$, close to the second-order transition at $\tau=2$.
The latter, indicated by the dashed vertical line, actually occurs
between $H_n$ and a homeotropic negative biaxial state $H_n^\dag$. At
the maximum of biaxiality (dotted vertical line), the state $H_n^\dag$
transforms continuously into the state $P^\dag$,
the two phases having the same symmetry.} \label{fig:phase}
\end{figure}

In fig.~\ref{fig:phase}, we show the typical surface phase diagrams in
the plane $(\tau,w)$ for two different values of $s_0$. At the bulk
phase transition $\tau=1$, for small values of $w$ there is always a
first-order transition between the states $P^\dag$ and $H_n$. For larger
values of $w$, two possible scenarios occur, depending on the value of
$s_0$ with respect to the bulk value at the transition
$s_b(1)=\sqrt{3/2}\simeq1.22$. When $s_0<s_b(1)$ (fig.~\ref{fig:phase}a)
there is a first-order transition between two different $P^\dag$ states.
When $s_0>s_b(1)$ (fig.~\ref{fig:phase}b), no surface transition occurs:
approaching $\tau=1$ from above, a wetting nematic layer of diverging
thickness continuously invades the whole sample. This corresponds to a
complete wetting situation, similar to the uniaxial case described in
Ref.~\cite{PingSheng,Miyano79,Tarczon80}; indeed, the surface-induced
biaxiality relaxes over $\simeq\xi$. In the isotropic phase ($\tau>1$),
a surface phase transition occurs at a given value of $w$. For
$s_0>s_b$ and $\tau<\tau_c$ (corresponding to the tricritical point in
fig.~\ref{fig:phase}b), the transition is of first order, between
the states $H_n$ and $P^\dag$. For $s_0<s_b$
(fig.~\ref{fig:phase}a), and for $\tau>\tau_c$ when $s_0>s_b$
(fig.~\ref{fig:phase}b), the transition is of second order, between
$H_n$ and a biaxial homeotropic negative state, $H_n^\dag$. In fact, by
symmetry, it is impossible to go continuously from $H_n$ to~$P^\dag$,
since, when the nematic director switches from one eigenvector to
another, the amount of biaxiality must be maximum [the order-parameter
tensor is then of the form~$\mathrm{diag}(0,\lambda,-\lambda)$]. This
behavior is illustrated in the inset of fig.~\ref{fig:phase}b. Note that
the transformation from $H_n^\dag$ to~$P^\dag$ is a continuous evolution
between two states having the same symmetry and not a phase transition:
the discontinuity in~$s_s$ and the cusp in~$B_s$, which occur when the
biaxiality is maximum, do not correspond to any actual discontinuity
of~$\mathsf{q}$. In fact, close to this transformation, the
tensor~$\mathsf{q}(z)$ at the surface~$z=0$ can be written as
\begin{equation}
\mathsf{q}(0) = \pmatrix{
\epsilon&0&0\cr
0&\sigma-\epsilon&0\cr
0&0&-\sigma},
\end{equation}
with~$\sigma>0$ and $\epsilon$ a small quantity ($|\epsilon|<\sigma$)
continuosly varying from positive to negative across the transformation.
In the~$H_n^\dag$ state, $\epsilon>0$: the eigenvalue of $\mathsf{q}(0)$
of maximum modulus is~$-\sigma<0$. This corresponds to a nematic
director along~$z$ (homeotropic ordering), with the \textit{negative}
scalar order-parameter~$s_s=-3\sigma/2$ and the amount of
biaxiality~$B_s=1-2\epsilon/\sigma$. In the ~$P^\dag$ state,
$\epsilon<0$: the eigenvalue of maximum modulus is
now~$\sigma-\epsilon>0$. This corresponds to a nematic director
along~$y$ (planar ordering), with the \textit{positive} scalar order
parameter~$s_s=3(\sigma-\epsilon)/2$ and the amount of
biaxiality~$B_s=1+2\epsilon/(\sigma-\epsilon)$. At the transformation,
$\epsilon=0$: no nematic director can be defined and the amount of
biaxiality is maximum ($B_s=1$). Therefore, the discontinuous change of
sign of~$s_s$ and the cusp of~$B_s$ at the transformation are not due to
a discontinuity of~$\mathsf{q}(0)$ but to the change of the direction
associated to the eigenvalue of maximum modulus. Note also that the
width of the~$H_n^\dag$ region is extremely narrow and goes to zero at
the tricritical point.

Our findings are compatible with the experimental data of Tarczon and
Miyano~\cite{Tarczon80}, who observed a pretransitional uniaxial
negative birefringence with optical axis perpendicular to the surface,
corresponding to our $H_n$ state, for the compound MBBA in contact with
a silane-treated substrate giving planar degenerate anchoring
in the nematic phase.  This \textit{negative} pretransitional
birefringence was not observed for the compound 5CB on the same
substrate, which probably corresponds to our $P^\dag$ surface state.

\section{Conclusions} As we have shown, the surface free energy
describing a \textit{planar degenerate} anchoring in the framework of
the Landau-de Gennes formalism must be at least of fourth-order in the
tensorial order-parameter $\mathsf{Q}$. Here, we have proposed the
simplest expression having this property and we have investigated the
associated anchoring and wetting behavior. In the nematic phase, we find
the expected planar degenerate anchoring with, however, a small surface
biaxiality. In the isotropic phase, a surface wetting layer appears: for
large enough surface couplings it is planar biaxial, while for low
couplings it is homeotropic uniaxial with negative scalar
order-parameter. These predictions are compatible with the experimental
observations for substrates yielding planar degenerate
anchoring~\cite{Tarczon80}. The transition between the two surface
states may be of first or second order. The biaxiality in the planar
states tends to zero for large surface coupling, yielding a uniaxial
degenerate planar anchoring or wetting. It would be interesting to study
the consequences of this degenerate anchoring in more complicated
geometries such as, e.g., colloidal suspensions.

\section{Note added in proofs} After this work was completed, Prof.
Sluckin kindly pointed out to us that the surface potentialof ref. [8],
linear in the tensor order-parameter, gives a genuine planar degenerate
anchoring in the case of complete wetting. We checked numerically that
this is true, although the ordering is strongly biaxial in a coherence
length close to the surface. Accordingly, above the transition
temperature, where the wetting layer is of thickness comparable to the
coherence length, the surface layer is strongly biaxial. Our
fourth-order potential, on the other hand, gives an almost perfectly
planar ordering in the whole microscopic layer, since it directly favors
the planar anchoring.

\acknowledgments We thank B. Silva for a critical reading of the
manuscript and useful comments.


\begin{thebibliography}{10}
\bibitem{det}Degenerate minima are possible if $W_{21}=0$, or if
$W_{21}=-\frac{2}{3}(W_{22}+W_{23})$ and $W_{11}=0$. The first case
corresponds to any $\mathsf{Q}$ possessing one eigenvector along
$\bm{\nu}$ (hence with arbitrary biaxiality). The second case corresponds
to $\mathsf{Q}$ uniaxial homeotropic with arbitrary $S$.
\bibitem{dem}According to the first condition, the non-zero components
of $\tQ_{ij}$, in the basis where the $z$-axis is along $\bm{\nu}$, can
be written as $\tQ_{xx}=\lambda$, $\tQ_{xy}=\tQ_{yx}=\mu$ and
$\tQ_{yy}=S_0-\lambda$. Then the second condition implies
$\mu^2=\lambda(S_0-\lambda)\ge0$.  Hence, there exists an angle $\theta$
such that $\lambda=S_0\cos^2\theta$ and $\mu=S_0\sin\theta\cos\theta$.
Thus, $\tQ_{ij}=S_0\,n_i\,n_j$ with $\bm{n}=(\cos\theta,\sin\theta,0)$.
\end{thebibliography}
\end{document}